# Ontology-Based Query Expansion with Latently Related Named Entities for Semantic Text Search

Vuong M. Ngo and Tru H. Cao

Faculty of Computer Science and Engineering
Ho Chi Minh City University of Technology
Viet Nam
{vuong.cs@gmail.com}
Pre-Print: 2010

**Abstract.** Traditional information retrieval systems represent documents and queries by keyword sets. However, the content of a document or a query is mainly defined by both keywords and named entities occurring in it. Named entities have ontological features, namely, their aliases, classes, and identifiers, which are hidden from their textual appearance. Besides, the meaning of a query may imply latent named entities that are related to the apparent ones in the query. We propose an ontology-based generalized vector space model to semantic text search. It exploits ontological features of named entities and their latently related ones to reveal the semantics of documents and queries. We also propose a framework to combine different ontologies to take their complementary advantages for semantic annotation and searching. Experiments on a benchmark dataset show better search quality of our model to other ones.

## 1 Introduction

With the explosion of information on the Word Wide Web and the emergence of e-societies where electronic documents are key means for information exchange, Information Retrieval (IR) keeps attracting much research effort, social and industrial interests. There are two types of searches in IR:

1. Document Retrieval: A user provides a search engine with a word, a phrase or a sentence to look for desired documents. The answer documents do not need to contain the terms in the user's query and can be ranked by their relatedness to the query. This type of searching was mentioned as Navigational Search in [14].
2. Question and Answering: A user provides a search engine with a phrase or sentence to look for objects, rather than documents, as answers for the user's query. This type of searching was mentioned as Research Search in [14].

In practice, answer objects obtained from a Question and Answering search engine can be used to search further for documents about them ([11]). Our work here is about Document Retrieval that uses related objects in a query to direct searching.

Current search engines like Yahoo and Google mainly use keywords to search for documents. Much semantic information of documents or user's queries is lost when they are represented by only 'bags of words'. Meanwhile, people often use named entities (NE) in information search. Specifically, in the top 10 search terms by YahooSearch[1] and GoogleSearch[2] in 2008, there are respectively 10 and 9 ones that are NEs. Named entities are those that are referred to by names such as people,

---

[1] http://buzz.yahoo.com/yearinreview2008/top10/
[2] http://www.google.com/intl/en/press/zeitgeist2008/

organizations, and locations ([22]) and could be described in ontologies.

The precision and recall measures of an IR system could be improved by exploiting ontologies. For Question and Answering, [17] proposed methods to choose a suitable ontology among different available ontologies and choose an answer in case of having various answers from different ontologies. Meanwhile, [25] presented a method to translate a keyword-based query into a description logic query, exploiting links between entities in the query. In [6], the targeted problem was to search for named entities of specified classes associated with keywords in a query, i.e., considering only entity classes for searching. Recently, in [15] the query was converted into SPARQL (a W3C standard language for querying RDF data) and the results were ranked by a statistical language model.

For Document Retrieval, the methods in [2] and [20] combined keywords with only NE classes, not considering other features of named entities and combinations of those features. In [5] and [11], a linear combination of keywords and NEs was applied, but a query had to be posted in RDQL to find satisfying NEs before the query vector could be constructed. Meanwhile, [12] proposed to enhance the content description of a document by adding those entity names and keywords in other documents that co-occurred with the entity names or keywords in that document. In [16], it was showed that normalization of entity names improved retrieval quality, which is actually what we call aliases here. As other alternative approaches, [26] and [9] respectively employed Wordnet and Wikipedia to expand query with related terms.

Our motivation and focus in this work is to expand a query with the names entities that are implied by, or related to, those in the query, which were not discovered in previous works. For example, given the query to search for documents about "*earthquakes in Southeast Asia*", documents about earthquakes in *Indonesia* or *Philippines* are truly relevant answers, because the two countries are part of *Southeast Asia*. Such named entities having relations with ones in a query are defined in an ontology being used. Intuitively, adding correct related named entities to a query should increase the recall while not sacrificing the precision of searching.

In this paper, we propose a new ontology-based IR model with two key ideas. First, the system extracts latently related named entities from a query to expand it. Second, it exploits multiple ontologies to have rich sources of both NE descriptions and NE relations for semantic expansions of documents and queries. Section 2 introduces the generalized Vector Space Model adapted from [4] combining keywords with different ontological features of named entities, namely, name, class, identifier, alias, and super-class. Section 3 describes the proposed system architecture and the methods to extract related named entities and to expand documents and queries. Section 4 presents evaluation of the proposed model and discussion on experiment results in comparison to other models. Finally, section 5 gives some concluding remarks and suggests future works.

## 2   A Generalized Vector Space Model

Textual corpora, such as web pages and blogs, often contain named entities, which are widely used in information extraction, question answering, natural language processing, and mentioned at Message Understanding Conferences (MUC) in 1990s

([19]). For example, consider the following passage from BBC-News³ written on Friday, 19 December 2008:

"The US government has said it will provide $17.4bn (£11.6bn) in loans to help troubled carmakers General Motors and Chrysler survive.

[...]

GM Chief Executive Rick Wagoner said his company would focus on: fully and rapidly implementing the restructuring plan that we reviewed with Congress earlier this month."

Here, *US*, *General Motors*, *Chrysler*, *GM* and *Rick Wagoner* are named entities.

Each NE may be annotated with its occurring name, type, and identifier if existing in the ontology of discourse. That is, a fully recognized named entity has three features, namely, name, type, and identifier. For instance, a possible full annotation of *General Motors* is the NE triple ("*General Motors*", *Company*, *#Company_123*), where *GM* and *General Motors* are aliases of the same entity whose identifier is *#Company_123*. Due to ambiguity in a context or performance of a recognition method, a named entity may not be fully annotated or may have multiple annotations. For instance, *Rick Wagoner* should be recognized as a person, though not existing in the ontology, hence its identifier is unknown.

As a popular IR model, the Vector Space Model (VSM) has advantages as being simple, fast, and with a ranking method as good as large variety of alternatives ([1]). However, with general disadvantages of the keyword based IR, the keyword based VSM is not adequate to represent the semantics of queries referring to named entities, for instances: (1) Search for documents about commercial organizations; (2) Search for documents about Saigon; (3) Search for documents about Paris City; (4) Search for documents about Paris City, Texas, USA.

In fact, the first query searches for documents containing named entities of the class *Commercial Organization*, e.g. NIKE, SONY, …, rather than those containing the keywords "commercial organization". For the second query, target documents may mention Saigon City under other names, i.e., the city's aliases, such as *Ho Chi Minh City* or *HCM City*. Besides, documents containing Saigon River or Saigon University are also suitable. In the third query, users do not expect to receive answer documents about entities that are also named "Paris", e.g. the actress *Paris Hilton*, but are not cities. Meanwhile, the fourth query requests documents about a precisely identified named entity, i.e., the Paris City in Texas, USA, not the one in France.

Nevertheless, in many cases, named entities alone do not represent fully the contents of a document or a query. For example, given the query "*earthquake in Indonesia*", the keyword "*earthquake*" also conveys important information for searching suitable documents. Besides, there are queries without named entities. Hence, it needs to have an IR model that combines named entities and keywords to improve search quality.

In [4], a generalized VSM was proposed so that a document or a query was represented by a vector over a space of generalized terms each of which was either a keyword or an NE triple. As usual, similarity of a document and a query was defined by the cosine of the angle between their representing vectors. The work implemented the model by developing a platform called S-Lucene modified from Lucene[4]. The system automatically processed documents for NE-keyword-based searching in the following steps:
1. Removing stop-words in the documents.
2. Recognizing and annotating named entities in the documents using KIM[5].
3. Extending the documents with implied NE triples. That is, for each entity named *n* possibly with class *c* and identifier *id* in the document, the triples (*n*/\*/\*), (\*/*c*/\*), (*n*/*c*/\*), (*alias*(*n*)/\*/\*), (\*/*super*(*c*)/\*), (*n*/*super*(*c*)/\*), (*alias*(*n*)/*c*/\*), (*alias*(*n*)/ *super*(*c*)/\*), and (\*/\*/*id*) were added for the document.
4. Indexing NE triples and keywords by S-Lucene.

---

³ http://news.bbc.co.uk/

Here *alias*(*n*) and *super*(*c*) respectively denote any alias of *n* and any super class of *c* in the ontology and knowledge base of discourse.

A query was also automatically processed in the following steps:
1. Removing stop-words in the query.
2. Recognizing and annotating named entities in the query.
3. Representing each recognized entity named *n* possibly with class *c* and identifier *id* by the most specific and available triple among (*n*/\*/\*), (\*/*c*/\*), (*n*/*c*/\*), and (\*/\*/*id*).

**Table 4.1.** Mapping interrogative words to entity types

| Interrogative Word | NE Class | Example Query |
|---|---|---|
| Who | Person | Who was the first American in space? |
|  | Woman | Who was the lead actress in the movie "Sleepless in Seattle"? |
| Which | Person | Which former Ku Klux Klan member won an elected office in the U.S.? |
|  | City | Which city has the oldest relationship as a sister-city with Los Angeles? |
| Where | Location | Where did Dylan Thomas die? |
|  | WaterRegion | Where is it planned to berth the merchant ship, Lane Victory, which Merchant Marine veterans are converting into a floating museum? |
| What | CountryCapital | What is the capital of Congo? |
|  | Percent | What is the legal blood alcohol limit for the state of California? |
|  | Money | What was the monetary value of the Nobel Peace Prize in 1989? |
|  | Person | What two researchers discovered the double-helix structure of DNA in 1953? |
| When | DayTime | When did the Jurassic Period end? |
| How | Money | How much could you rent a Volkswagen bug for in 1966? |

However, [4] did not consider latent information of the interrogative words *Who*,

---
[4] http://lucene.apache.org/
[5] http://www.ontotext.com/kim/

*What*, *Which*, *When*, *Where*, or *How* in a query. For example, given the query "*Where was George Washington born?*", the important terms are not only the NE *George Washington* and the keyword "*born*", but also the interrogative word *Where*, which is to search for locations or documents mentioning them. The experiments on a TREC dataset in [21] showed that mapping such interrogative words to appropriate NE classes improved the search performance. For instance, *Where* in this example was mapped to the class *Location*. The mapping could be automatically done with high accuracy using the method proposed in [3]. Table 4.1 gives some examples on mapping interrogative words to entity types, which are dependent on a query context.

## 3 Ontology-Based Query Expansion

### 3.1 System Architecture

Our proposed system architecture of semantic text search is shown in Figure 3.1. It has two main parts. Part 1 presents the generalized VSM searching system implemented in [21]. Part 2 presents the query expansion module, which is the focus of this paper, to add in a query implied, i.e., latently related, named entities before searching.

The NE Recognition and Annotation module extracts and embeds NE triples in a raw text. The text is then indexed by contained NE triples and keywords and stored in the Extended NE-Keyword-Annotated Text Repository. Meanwhile, the InterrogativeWord-NE Recognition and Annotation module extracts and embeds the most specific NE triples in the extended query and replaces the interrogative word if existing by a suitable class. Semantic document search is performed via the NE-Keyword-Based Generalized VSM module.

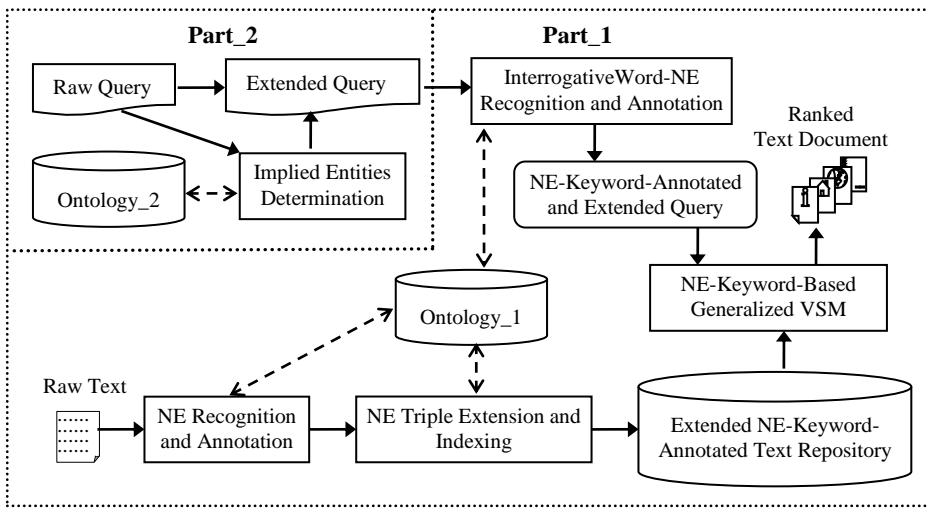

**Fig. 3.1.** System architecture for semantic text search

An ontology is a formal description of classes, entities, and their relations that are assumed to exist in a world of discourse ([13], [10]). Since no single ontology is rich enough for every domain and application, merging or combining multiple ontologies are reasonable solutions ([7]). Specifically, on one hand, our proposed model needs an ontology with a comprehensive class catalog, large entity population and rich entity description, and an efficient accompanying NE recognition engine, for annotating documents and queries. On the other hand, it needs one with many relations between entities, for expanding queries with latently related named entities.

In this work we employ KIM ([18]) for Ontology_1 in the system architecture illustrated above, as an infrastructure for automatic NE recognition and semantic annotation of documents and queries. The used KIM ontology is an upper-level ontology containing about 250 concepts and 100 attributes and relations. KIM Knowledge Base (KB) contains about 77,500 entities with more than 110,000 aliases. NE descriptions are stored in an RDF(S) repository. Each entity has information about its specific type, aliases, and attributes (i.e., its own properties or relations with other named entities).

However, KIM ontology defines only a small number of relations. Therefore, we employ YAGO (Yet Another Great Ontology) ([23], [24]), which is rich in assertions of relations between named entities, for Ontology_2 in the system. It contains about 1.95 millions entities, 93 different relation types, and 19 millions facts that are specific relations between entities. The facts are extracted from Wikipedia and combined with WordNet using information extraction rules and heuristics. New facts are verified and added to the knowledge base by YAGO core checker. Therefore the correctness of the facts is about 95%. In addition, with logical extraction techniques and a flexible architecture, YAGO can be further extended in future. Note that, to have more relation types and facts, we can employ and combine it with some other ontologies.

### 3.2 Query Expansion

Figure 3.2 shows the main steps of our method to determine latently related entities for a query:

1. Recognizing Relation Phrases: Relation phrases are prepositions, verbs, and other phrases representing relations, such as *in*, *on*, *of*, *has*, *is*, *are*, *live in*, *located in*, *was actress in*, *is author of*, *was born*. We implement relation phrase recognition using the ANNIE tool of GATE ([8]).
2. Determining Relations: Each relation phrase recognized in step 1 is mapped to a corresponding one in Ontology_2 by a manually built dictionary. For example, "*was actress in*" is mapped to *actedIn*, "*is author of*" is mapped to *wrote*, and "*nationality is*" is mapped to *isCitizenOf*.
3. Recognizing Entities: Entity recognition is implemented by OCAT (Ontology-based Corpus Annotation Tool) of GATE.
4. Determining Related Entities: Each entity that has a relation determined in step 2 with an entity recognized in step 3 is added to the query. In the scope of this paper, we consider to expand only queries having one relation each. However, the method can be applied straightforwardly to queries with more than one relation.

After the query is expanded with the names of latently related entities, it is processed by Part 1 of the system described in above.

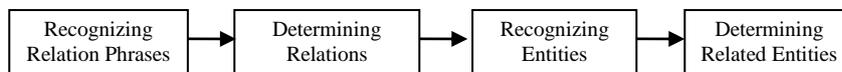

**Fig. 3.2.** The steps determining latently related entities for a query

## 4 Experiment

### 4.1 Datasets

A test collection includes 3 parts: (1) document collection; (2) query collection; and (3) relevance evaluation, stating which document is relevant to a query. A document is relevant to a query if it actually conveys enquired information, rather than just the words in the query. There are well-known standard datasets such as TREC, CISI, NTCIR, CLEF, Reuters-21578, TIME, and WBR99.

**Table 4.2.** Statistics about dataset usage of text retrieval papers in SIGIR 2007 and SIGIR 2008

| SIGIR | Paper Total | Number of Papers Using a Dataset Type | | |
|---|---|---|---|---|
| | | Author-Own Dataset | Other Standard Dataset | TREC's Dataset |
| **2007** | 34 | 11 | 7 | 21 |
| **2008** | 22 | 8 | 4 | 12 |
| **2007+2008** | 56 | 19 (~34%) | 11 (~20%) | 33 (~59%) |

We have surveyed papers in SIGIR-2007[6] and SIGIR-2008[7] to know which datasets have been often used in information retrieval community so far. We only consider papers about text IR, not IR for multi-language, picture, music, video, markup document XML, SGML,… Besides, all poster papers in SIGIR-2007 and SIGIR-2008 are not reviewed. There are 56 papers about text IR examined and classified into three groups, namely, TREC[8] (The Text REtrieval Conference), author-own, and other standard datasets. TREC is annually co-organized by the National Institute of Standards and Technology (NIST) and U.S. Department of Defense, supporting research and evaluation of large-scale information retrieval systems. Table 4.2 shows that 59% of papers use TREC's datasets as popular ones in the IR community.

We choose the L.A. Times document collection, which was a TREC one used by 15 papers among the 33 papers of SIGIR-2007 and SIGIR-2008 in the above survey. The L.A. Times consists of more than 130,000 documents in nearly 500MB. Next, we choose 124 queries out of 200 queries in the QA Track-1999 that have answer documents in this document collection.

### 4.2 Testing results

Using the chosen dataset, we evaluate the performance of the proposed model and compare it with others by the common precision (P) and recall (R) measures ([1]). In addition, our system ranks documents regarding their similarity degrees to the query.

---

[6] http://www.sigir2007.org

[7] http://www.sigir2008.org

[8] http://trec.nist.gov

Hence, P-R curves represent better the retrieval performance and allow comparing those of different systems. The closer the curve is to the right top corner, the better performance it represents.

In order to have an average P-R curve of all queries, the P-R curve of each query is interpolated to the eleven standard recall levels 0%, 10%, …, 100% ([1]). Besides, a single measure combining the P and R ones is F-measure, which is computed by $F = \frac{2.P.R}{P+R}$. We also use average F-R curves at the eleven standard recall levels to compare system performances.

We conduct experiments to compare the results obtained by three different search models:
1. Keyword Search: This search uses Lucene text search engine.
2. NE+KW Search: This search is given in [21].
3. Semantic Search: This is the search engine proposed in this paper.

**Table 4.3.** The average precisions and F-measures at the eleven standard recall levels on 124 queries of the L.A. Times

| Measure | Model | Recall (%) | | | | | | | | | | |
|---|---|---|---|---|---|---|---|---|---|---|---|---|
| | | 0 | 10 | 20 | 30 | 40 | 50 | 60 | 70 | 80 | 90 | 100 |
| Precision (%) | Lucene | 66.1 | 66.0 | 63.2 | 60.4 | 56.7 | 55.1 | 45.7 | 40.4 | 37.9 | 37.5 | 37.1 |
| | NE+KW | 71.8 | 71.6 | 69.5 | 65.5 | 62.2 | 60.8 | 52.4 | 48.0 | 46.4 | 45.4 | 44.7 |
| | Semantic Search | 73.0 | 72.7 | 70.9 | 67.0 | 63.8 | 62.4 | 54.4 | 50.1 | 48.4 | 47.4 | 46.8 |
| F-measure (%) | Lucene | 0.0 | 15.5 | 26.6 | 34.8 | 40.1 | 45.0 | 43.4 | 42.2 | 41.8 | 43.0 | 44.1 |
| | NE+KW | 0.0 | 16.3 | 28.4 | 37.1 | 42.8 | 48.3 | 48.0 | 47.7 | 48.5 | 49.8 | 50.8 |
| | Semantic Search | 0.0 | 16.5 | 28.7 | 37.4 | 43.3 | 49.0 | 48.8 | 48.7 | 49.6 | 51.0 | 52.1 |

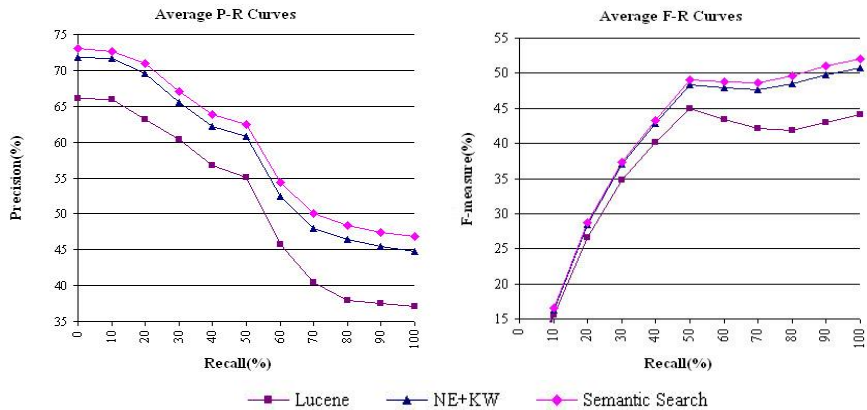

**Fig 4.1.** Average P-R and F-R curves of Lucene, KW+NE and Semantic Search models on 124 queries of the L.A. Times

From Table 4.3 and Figure 4.1, we can see the average precisions and F-measures of the keyword-based Lucene, the NE+KW Search, and the proposed Semantic Search at each of the standard recall levels of 124 queries. They show that taking into account latent ontological features in queries, documents, and expanding queries using relations described in an ontology enhance text retrieval performance.

In the 124 queries, there are only 17 queries expanded. The other queries are not expanded because: (1) The queries have more than one relation phrase, which are out of the experiment scope of this paper (55 queries); (2) Ontoloy_2 does not have relation types corresponding to the relation phrases in the queries (36 queries); and (3) Ontoloy_2 does not have facts asserting specific relations of named entities in the queries with others (16 queries). Table 4.4 and Figure 4.2 show the average precisions and F-measures of the three systems at each of the standard recall levels for those 17 expanded queries only. One can observe that, when all queries are expanded using our proposed method, the Semantic Search clearly outperforms the other two systems.

**Table 4.4.** The average precisions and F-measures at the eleven standard recall levels on 17 expanded queries of the L.A. Times

| Measure | Model | Recall (%) | | | | | | | | | | |
|---|---|---|---|---|---|---|---|---|---|---|---|---|
| | | 0 | 10 | 20 | 30 | 40 | 50 | 60 | 70 | 80 | 90 | 100 |
| Precision (%) | Lucene | 61.9 | 61.9 | 58.4 | 53.6 | 51.9 | 51.9 | 39.6 | 39.1 | 38.3 | 38 | 37.6 |
| | NE+KW | 71.6 | 71.6 | 69.5 | 67.7 | 66.9 | 65.8 | 55.2 | 54.9 | 54.8 | 54.7 | 54.7 |
| | Semantic Search | 82.2 | 82.2 | 82.2 | 80.1 | 80.1 | 78.9 | 70.3 | 70.3 | 69.7 | 69.7 | 69.7 |
| F-measure (%) | Lucene | 0.0 | 14.7 | 25.0 | 32.3 | 37.4 | 42.5 | 38.3 | 40.4 | 41.6 | 43.2 | 44.2 |
| | NE+KW | 0.0 | 16.3 | 28.5 | 37.6 | 44.8 | 51.2 | 50.0 | 52.8 | 55.3 | 57.9 | 58.9 |
| | Semantic Search | 0.0 | 17.5 | 31.1 | 40.4 | 48.7 | 56.0 | 55.8 | 59.6 | 62.6 | 66.0 | 67.5 |

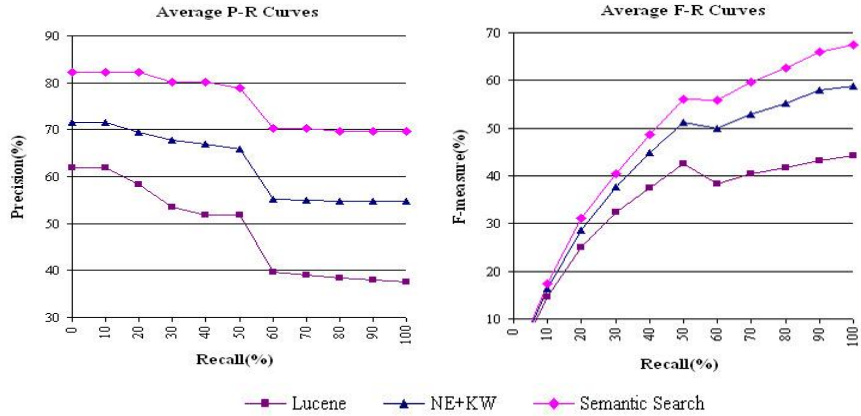

**Fig 4.2.** Average P-R and F-R curves of Lucene, KW+NE and Semantic Search models on 17 expanded queries of the L.A. Times

We have analyzed some typical queries for which Semantic Search is better or worse than NE+KW. For query_38 "*Where was George Washington born*?", Semantic Search performs better than NE+KW. Actually, Semantic Search maps the relation word *born* to the relation *bornIn* and Ontology_2 has the fact (*George_Washington bornIn Westmoreland_Country*). So, *Westmoreland Country* is added to the query.

For query_190 "*Where is South Bend?*", Semantic Search maps the relation phrase *where is* to the relation *locatedIn*, and Ontology_2 has the fact (*South_Bend locatedIn Indiana*). However, all of the relevant documents of the query only contain *Ind* rather than *Indiana*. Although, *Ind* is an alias of *Indiana*, Ontology_1 does not include it. Therefore, when adding *Indiana* into the query, it makes Semantic Search perform worse than NE+KW.

## 5   Conclusion and Future Works

We have presented the generalized VSM that exploits ontological features for semantic text search. That is a whole IR process, from a natural language query to a set of ranked documents. Given an ontology, we have explored latent named entities related to those in a query and enriched the query with them. We have also proposed a framework to combine multiple ontologies to take their complementary advantages for the whole semantic search process.

Besides, the system takes into account all the main features of named entities, namely, name, class, identifier, alias, super-class, and supports various query types, such as searching by only named entities, only keywords, combined named entities and keywords, and *Wh*-questions. The conducted experiments on a TREC dataset have showed that appropriate ontology exploitation improves the search quality in terms of the precision, recall, and F-measures. In particular, expanding queries with implied named entities, the proposed Semantic Search system outperforms the previous NE+KW and keyword-based Lucene ones.

Our experiments on the effect of query expansion are partially eclipsed by the used Ontology_2, which does not cover all relation types and specific relations in the test query set, and by the relation recognition module of our system. For future work, we will combine more ontologies to increase the relation coverage and research logical methods to recognize better relations in a query. Furthermore, we will also investigate ontological similarity and relatedness between keywords. These are expected to increase the performance of the proposed model.